\documentstyle[preprint,aps]{revtex}
\begin{document}
\draft
%
%
\title{Ni-Au:  A testing ground for theories of phase stability}
\author{C. Wolverton and Alex Zunger}
\address{National Renewable Energy Laboratory, Golden, CO 80401\\}
\date{\today}
\maketitle
\begin{abstract}
The theory of
phase stability in the  Ni-Au alloy system is a popular topic
due to the large size mismatch between Ni and Au,
which makes the effects of atomic relaxation critical,
and also the fact that Ni-Au exhibits a phase separation 
tendency at low temperatures, but measurements at
high-temperature show an ordering-type short-range order.
We have clarified the wide disparity which exists in
the previously calculated values of mixing energies 
and thermodynamic properties
by computing ``state-of-the-art'' energetics 
(full-potential, fully-relaxed LDA total energies)
combined with ``state-of-the-art'' statistics
({\bf k}-space cluster expansion with Monte Carlo simulations)
for the Ni-Au system.  We find:
(i) LDA provides accurate mixing energies of 
disordered Ni$_{1-x}$Au$_x$ alloys ($\Delta H_{\rm mix}\lesssim +100$ meV/atom)
provided that both
atomic relaxation (a $\sim$100 meV/atom effect) and
short-range order ($\sim$25 meV/atom) are taken into
account properly.
(ii) Previous studies using empirical potentials or
approximated LDA methods often underestimate the formation
energy of ordered compounds, and hence also underestimate
the mixing energy of random alloys.
(iii) Measured values of the total entropy of mixing combined
with calculated values of the configurational entropy
demonstrate that the {\em non-configurational} entropy in
Ni-Au is large, and leads to a significant reduction in
miscibillity gap temperature. 
(iv) The calculated short-range order agrees well with
measurements, and both predict {\em ordering} in the
disordered phase.  
(v) Consequently, using inverse Monte Carlo to
extract interaction energies from the
measured/calculated short-range order in Ni-Au
would result in interactions which would produce
ordering-type mixing energies, in contradiction 
with both experimental measurements and precise LDA 
calculations.
\end{abstract}

\eject

\section{Introduction}

The Ni-Au alloy system is physically interesting because, on one hand
it exhibits a phase separation tendency at low
temperatures and {\em positive} mixing enthalpies \cite{Hultgren} and,
on the other hand, an ordering-type short-range order (SRO) 
at high temperatures. \cite{Cohen2}
Also, the fcc Ni and Au constituents possess a large lattice-mismatch 
($\Delta a/a \sim 15$\%), thus making this system a critical
test for any alloy phase stability theory hoping to capture
the effects of atomic relaxation.
Important early experimental and theoretical work on this alloy includes
the work of Moss {\em et al.} \cite{Moss1,Moss2},
Cohen {\em et al.} \cite{Cohen1,Cohen2,Cohen3}, and
and Cook and de Fontaine \cite{Cook}.  
The coexistence of phase separation (at low $T$) with short-range ordering 
(at high $T$) in the
same alloy system might have been naively construed to imply a change from
repulsive (``ferromagnetic'') interactions at low $T$ to attractive
(``anti-ferromagnetic'') interactions at higher $T$.  The change would
have been surprising, given that no electronic, magnetic, or structural
change is observed in this temperature range.  The answer to this puzzle
was given by Lu and Zunger: \cite{Lu94}
The excess energy for a disordered Ni$_{1-x}$Au$_x$
alloy or an ordered compound of type $\sigma$ is given by:
\begin{equation}
\Delta H_{\sigma} = E_{\sigma}(a_{\rm eq}^{\sigma})
- [(1-x)E_{\rm Ni}(a_{\rm eq}^{\rm Ni})
+ xE_{\rm Au}(a_{\rm eq}^{\rm Au})],
\end{equation}
and may be written \cite{e-G}
$\Delta H = \Delta \epsilon + \Delta E_{\rm VD}$, where
$\Delta \epsilon$ is the constant-volume, ``spin-flip''
energy required to create $\sigma$ out
of Ni and Au, each already
prepared at the alloy lattice constant $a_{\rm eq}^{\sigma}$,
and $\Delta E_{\rm VD}$ is the volume deformation 
energy required to hydrostatically
deform Ni and Au from their respective
$a_{\rm eq}^{\rm Ni}$ and $a_{\rm eq}^{\rm Au}$
to $a_{\rm eq}^{\sigma}$.
In Ref. \cite{Lu94}, it is demonstrated that SRO is
determined by the constant volume energy change
$\Delta \epsilon$, which is negative (ordering, or ``anti-ferromagnetic'')
in Ni-Au, indicating
an ordering tendency of SRO.  However, 
$\Delta E_{\rm VD} \equiv G(x)$
is large and positive, making $\Delta H > 0$.  And, since
long-range order is determined by $\Delta H$, Ni-Au shows phase-separating
(``ferromagnetic'') long-range order.
This analysis leads to two unexpected conclusions:  First, that the
time-honored Ising-like representation of alloy thermodynamics which
includes only ``spin-flip'' energies of the $\Delta \epsilon$ type,
but ignores the elastic energy $G(x)$ will fail in explaining basic
stability trends for systems such as 
Ni-Au.  Second, since
measurements or calculations of the SRO
are insensitive to physical effects (i.e., elastic deformation $\Delta E_{\rm VD}$)
that control measurements/calculations of mixing enthalpies
$\Delta H$, the 
often-used practice \cite{imc} of ``inverting'' the SRO profile to extract
interaction energies that are then used to predict mixing enthalpies
is {\em fundamentally flawed}.  Specifically, inversion of the
SRO of Ni-Au will produce ordering-like interaction energies which,
when used to calculate mixing enthalpies will produce (ordering-like)
negative values, while the measured ones are strongly positive. \cite{Hultgren,emf} 

For these and other reasons,
the theory of phase stability in Ni-Au has recently
become quite popular 
\cite{Lu94,Pasturel,Mohri,Amador,Colinet,Morgan,Eymery,Finel,Asta}
(Table \ref{comp.niau}). 
These calculations are distinguished by the methods used for
(i) energetics ($T$=0 K) and (ii) statistics ($T\neq 0$).
Energy calculations ($T$=0 K) 
for this system have been performed by a wide variety of
techniques:  First-principles calculations, both full-potential
(FLAPW \cite{Lu94} and FLMTO\cite{Pasturel}) and 
atomic-sphere-approximation (LMTO\cite{Amador,Colinet,Morgan} and
ASW\cite{Mohri}), as well as 
semi-empirical (EAM\cite{Asta}) and empirical potentials 
\cite{Finel,Pasturel,Eymery}.
There are significant variations in the computed energetics (Table \ref{comp.niau}).
Statistics have been applied to these calculations using
cluster expansions (CE) such as $\epsilon$-G \cite{e-G}, 
Connolly-Williams \cite{Connolly83}, and second-order expansions. \cite{deFontaine79}

The purpose of this paper is thus three-fold:

{\em First}, we would like to clarify the conflicting energetic and
statistical results (Table \ref{comp.niau}) by computing
``state-of-the-art'' energetics for Ni-Au alloys (full-potential
LAPW total energies including full atomic relaxation) combined with
``state-of-the-art'' statistics (a k-space CE \cite{Laks92}
with Monte Carlo simulations).  
These computations will clarify whether the better
agreement with experimental $\Delta H$ obtained by approximated methods
(e.g., empirical and semi-empirical potentials, as well as atomic-sphere-approximation
methods) 
relative to full LDA methods is fundamental or accidental.

{\em Second}, we would like to address the issue of why the calculated 
miscibility gap temperatures are often much too high compared with
the experimentally assessed phase diagram \cite{Hultgren}.
In Table \ref{comp.niau}, one can see a fixed ratio between
calculated miscibility gap temperatures $T_{\rm MG}$ 
and the calculated $\Delta H_{\rm mix}$.  In fact, all previous
calculations (except the EAM calculations of Asta and Foiles \cite{Asta})
very nearly follow the ratio obtained using mean-field configurational
entropy: $k_BT_{\rm MG}/\Delta H_{\rm mix} = 2$.  However, the
{\em experimental} value of this ratio is 1.2.
We will examine this apparent discrepancy between experimental
$\Delta H_{\rm mix}$ and $T_{\rm MG}$ below.

{\em Third}, we would like to examine the SRO
in Ni-Au and discuss the implications of this SRO on 
``inverse'' techniques, mentioned above, for calculating phase stability in alloys.
We will offer a challenge to practitioners of the inverse Monte Carlo method.

\section{Checking Ordered Compound Formation Energies}

Table \ref{comp.niau} summarizes the previous calculations on
the mixing enthalpies of random
Ni-Au alloys.  The wide discrepancy between 
calculated values of $\Delta H_{\rm mix}$ (48-170 meV/atom)
is apparent from this table.  
Since mixing enthalpies $\Delta H_{\rm mix}$ of random alloys
can be expressed [see, e.g., Eq.\ (3b) in Ref. \onlinecite{Zunger94}]
as a linear combination of formation enthalpies $\Delta H_f(\sigma)$
of certain ordered compounds $\{\sigma\}$, the discrepancies
in $\Delta H_{\rm mix}$ must reflect discrepancies in $\Delta H_f(\sigma)$.
But formation enthalpies of small-unit-cell ordered compounds can be
computed accurately and reliably via full-potential fully-relaxed LDA
methods.  Our strategy will thus be to trace 
the source of the discrepancy in $\Delta H_{\rm mix}$
to the values of formation energies
of various Ni$_p$Au$_q$ ordered compounds, as shown in 
Table \ref{comp.niau.ord}.  Examining this table leads to
several interesting points regarding the energetics in Ni-Au:

\subsection{FLMTO vs. ASA methods (LMTO, ASW)}

In comparing the full-potential LMTO \cite{Pasturel} to
LMTO-ASA \cite{Morgan} calculations, one can see significant
and strongly configuration-dependent
discrepancies, {\em even when considering unrelaxed configurations}.
For example, the $Z2$ structure [a Ni$_2$Au$_2$ (001) superlattice]
has an unrelaxed formation energy which is nearly 100 meV/atom
lower in the LMTO-ASA calculation than in the full-potential LMTO one.
Thus, the ASA-based calculations (LMTO, ASW) in the Ni-Au system
cannot be trusted for the kind of quantitative
energetics required in phase stability studies. \cite{Wolverton94}

\subsection{Harmonic vs. anharmonic relaxation}

In a large lattice-mismatched system like Ni-Au,
the effects of atomic relaxation are likely to be crucial.
Although straightforward,
fully relaxing all of the cell-internal and cell-external
degrees of freedom can be computer intensive.  One alternative
to full atomic relaxation (using quantum mechanical forces and
total energy minimization)
which has been used in Ni-Au \cite{Lu94}
is to use continuum elasticity theory \cite{elastic}
to find the relaxed geometry, with a subsequent LDA calculation
with this geometry to find the relaxed energetics.
Continuum elasticity theory can be used as a relaxation
model by realizing that many ordering Ni$_p$Au$_q$
compounds can be described as ``superlattices'' along
some special orientations $\hat{k}$.  Continuum elasticity
then provides the equilibrium interlayer spacing $c_{\rm eq}$
along $\hat{k}$
as a function of the externally-fixed perpendicular lattice
constant $a_{\perp}$ as the minimum of the epitaxial
strain energy  due to the external constraint:
\begin{eqnarray}
\label{cont.elasticity}
c_{\rm eq}(\hat{k},a_{\perp})  &=& a_{\rm eq}^{(\lambda)} + 
      [2-3q^{(\lambda)}(a_{\perp},\hat{k}) 
      [a_{\rm eq}^{(\lambda)}-a_{\perp}] + ...  \nonumber \\
\\
q(a_{\perp},\hat{k}) &=& \frac{\Delta E_{\rm epi}^{\rm eq}(a_{\perp},\hat{k})}
		            {\Delta E_{\rm bulk}(a_{\perp})} 
\end{eqnarray}
where $E_{\rm eq}^{(\lambda)}$  and $a_{\rm eq}^{(\lambda)}$
are the equilibrium energy and lattice constant of the 
cubic material $\lambda$.  
$\Delta E_{\rm epi}^{\rm eq}$ is the energy
of the alloy constituent subject to the biaxial constraint
that the lattice constant perpendicular to $\hat{k}$ is 
externally fixed to be $a_{\perp}$.
$\Delta E_{\rm bulk}(a_{\perp})$ is simply the 
deformation energy change upon hydrostatically distorting the
material from $a_{\rm eq}$ to $a_{\perp}$.
The central quantity in these elastic calculations is 
the ``strain reduction factor'' 
$q(a_{\perp},\hat{k})$.
In continuum elasticity
theories, $q(a_{\perp},\hat{k})$
is given by
\begin{equation}
\label{q.harm}
q(a_{\perp},\hat{k}) = 1 - B/[C_{11}+\gamma(a_{\perp},\hat{k})\Delta].
\end{equation}
where
\begin{equation}
\label{delta}
\Delta = C_{44}-(C_{11}-C_{12})/2
\end{equation}
is the
elastic anisotropy, $B=(C_{11}+2C_{12})/3$ is the bulk modulus, and $C_{ij}$
are elastic constants.  In the {\em harmonic} approximation, 
$q(a_{\perp},\hat{k})$ is further assumed to be $a_{\perp}$-independent,
and $\gamma^{\rm harm.}(\hat{k})$ is the following
geometric function for the direction $\hat{k} = (l,m,n)$: 
\begin{eqnarray}
\label{gamma}
\gamma_{\rm harm.}(l,m,n) &=& 
\frac{4(l^2m^2 + m^2n^2 + n^2l^2)}{(l^2 + m^2 + n^2)^2} \nonumber \\
&=& \frac{4}{5}\sqrt{4\pi}\;[K_0(l,m,n) - \frac{2}{\sqrt{21}}K_4(l,m,n)]
\end{eqnarray}
where $K_L$ are the Kubic harmonics of angular momentum $L$.

Using Eqs.\ (\ref{cont.elasticity})-(\ref{gamma}) thus provides
predicted relaxed geometries $c_{\rm eq}(\hat{k},a_{\perp})$
for alloy compounds (e.g., the
$Z2$ structure) given the elastic constants and $a_{\rm eq}^{(\lambda)}$.  
Indeed, these equations
have been routinely used (see review in Ref. \onlinecite{Zungerepi})
to predict the distortion $c_{\rm eq}-a_{\rm eq}$ of films grown 
epitaxially on a substrate with lateral lattice constant $a_{\perp}$.
Comparison to LDA calculations \cite{Bernard94} shows that for semiconductors
with lattice mismatch $(a_{\rm eq}-a_{\perp})/a_{\perp} \lesssim 7\%$,
the harmonic expressions (\ref{q.harm})-(\ref{gamma}) work very well
down to a monolayer thickness.  However, we find that for noble- and
transition-metal alloys with a much larger lattice mismatch
(e.g., Ni-Au, Cu-Au with $\Delta a/a$ = 15\%, 12\%, respectively),
anharmonic corrections are important.  As we will see below in Sec. \ref{anharm.cs}, 
this is manifested by the fact that $\gamma(a_{\perp},\hat{k})$ 
of Eq. (\ref{q.harm})
now has additional
terms to those appearing in the harmonic form of Eq. (\ref{gamma}).  
These anharmonic terms in $\gamma(\hat{k})$ lead via Eq. (\ref{q.harm}) to 
corrections to $q(a_{\perp},\hat{k})$, and consequently via 
Eq. (\ref{cont.elasticity}) to the
relaxation of the lattice constant $c_{\rm eq}(\hat{k})$.
Indeed, using the same FLAPW as Ref. \onlinecite{Lu94}, 
but minimizing the total energy
quantum-mechanically 
(``Fully relaxed'' in Table \ref{comp.niau.ord})
rather than via the harmonic expression of 
Eq.\ (\ref{q.harm})
(``Partially relaxed'' in Table \ref{comp.niau.ord}), 
we find a lower-energy relaxation for $Z2$:
The LDA energy minimization gives $\Delta H(Z2)$ = +70.2
meV/atom,
while LDA with harmonic relaxation gives +124.3.
For other structures, the effect is much lower.  Nevertheless,
anharmonic relaxation in Ni-Au alloys 
is large and cannot be neglected.

\subsection{Empirical methods:  Getting the right $\Delta H_{\rm mix}(x,T)$
for the wrong reason?}  

We see from Table \ref{comp.niau} that
the methods that use empirical evaluations of $\Delta H_{\rm mix}(1/2,\infty)$
\cite{Morgan,Eymery,Finel,Pasturel,Asta}
produce results that are lower, and thus closer to the measured
$\Delta H_{\rm mix}(1/2,1150)$ than methods that use converged,
full potential, fully relaxed approaches (i.e., the present work and
Refs. \onlinecite{Lu94,Pasturel}). 
Since there is a 
proportionality between $\Delta H_{\rm mix}$ and $\Delta H_f(\sigma)$,
we thus surmise that the empirical methods will produce formation
energies $\Delta H_f(\sigma)$ of ordered compounds that are lower than
the LDA results for such systems.
Indeed, Table \ref{comp.niau.ord} shows the formation energies of
two of the empirical potential methods.  By comparing these
numbers to full-potential LDA energies, one can see that the
{\em empirical potentials systematically underestimate the formation
energies of ordered compounds}.
Since the LDA method is expected to reproduce formation enthalpies
of small-unit-cell ordered structures rather accurately, and since
FLAPW gives a precise representation of the LDA, we think that
the underestimation of FLAPW energies by the empirical methods
is a rather serious limitation of these methods.
The EAM of Ref. \onlinecite{Asta} was fit to the unrelaxed FLAPW
calculations of Ref. \onlinecite{Lu94}, and thus reproduces
these energies fairly well (except for the $Z2$ structure).
However, the EAM severely overestimates the energetic effect of 
relaxation, and hence produces relaxed formation energies which
are much lower than LDA, and in some cases are even negative. \cite{note.3}
It would be desirable to see more formation energies of ordered
compounds from the empirical methods to determine test the
expectation of underestimation of $\Delta H_f(\sigma)$ relative
to LDA.

In summary, the reason that empirical methods agree with measured
random-alloy mixing enthalpy better than LDA methods do is a
systematic underestimation by the empirical methods of even the
ordered compound energies.

%

\section{Present Calculations - FLAPW with k-space Cluster Expansion}

\subsection{FLAPW calculations of ordered compounds}

We have performed first-principles full-potential LAPW \cite{singh}
calculations for pure Ni, pure Au, and a large number (31) 
of fcc-based
Ni-Au compounds in order to construct an accurate
cluster expansion.  
The total energy of each compound is fully minimized 
with respect to volume, cell-internal, and 
cell-external \cite{note} coordinates using
quantum-mechanical forces.
We have used the exchange correlation of
Wigner \cite{wigner}.  The muffin-tin radii are chosen to be
2.2 a.u for Ni and 2.4 a.u. for Au.  Brillouin-zone integrations
are performed using the equivalent {\bf k}-point sampling
method, \cite{froyen}  with the {\bf k}-points for each
compound all mapping into the same 60 special {\bf k}-points
for the fcc structure.  This mapping guarantees that the
total energy per atom of an elemental metal calculated either with
the fcc unit cell or with a lower symmetry (e.g., any of the
compounds) are identical.  All calculations performed are
non-magnetic.  (The spin polarization energy difference
between ferro- and non-magnetic fcc Ni was calculated and
found to be -50 meV/atom.)

The 31 calculated LAPW formation energies 
are given in Table \ref{niau.eform}.
Both relaxed and unrelaxed (total energy minimized with respect to volume, 
but with cell-internal and cell-external coordinates held fixed at ideal fcc positions)
formation energies are shown.
The nomenclature of the compounds studied is the same
as given in \cite{Zunger94}.
Many of the compounds considered can be described
as Ni$_p$Au$_q$ ``superlattices'' along a particular
orientation $\hat{k}$:
\begin{eqnarray}
&&{\rm Ni}_1{\rm Au}_1: \; [100], [111], \nonumber \\
&&{\rm Ni}_2{\rm Au}_1: \; [100], [011], [111], \nonumber \\
&&{\rm Ni}_1{\rm Au}_2: \; [100], [011], [111], \nonumber \\
&&{\rm Ni}_3{\rm Au}_1: \; [100], [011], [201], [111], [311], \nonumber \\
&&{\rm Ni}_1{\rm Au}_3: \; [100], [011], [201], [111], [311], \nonumber \\
&&{\rm Ni}_2{\rm Au}_2: \; [100], [011], [201], [111], [311], \nonumber \\
&&{\rm Ni}_3{\rm Au}_3: \; [100],  \nonumber \\
&&{\rm Ni}_2{\rm Au}_3: \; [100]. 
\end{eqnarray}
We also calculated the energies of six other structures:
$L1_2$(Ni$_3$Au and NiAu$_3$),
$D7$(Ni$_7$Au and NiAu$_7$), and two
8-atom ``special quasi-random structures''\cite{SQS}, 
SQS14$_a$ (Ni$_6$Au$_2$ and Ni$_2$Au$_6$).
In addition the Ni$_p$Au$_q$ long-period superlattice limits
($p,q \rightarrow \infty$)
needed in the construction of the k-space cluster
expansion (see below) were computed for six principle directions:
[100], [011], [201], [111], [311], and [221].
The numerical error of the LAPW calculations of $\Delta H_f$ is
estimated to be $\sim$10 meV/atom or less.

\subsection{{\bf k}-space cluster expansion}

The Ni-Au formation energies $\Delta H_{\sigma}$
for structures $\sigma$ are then mapped 
onto a cluster expansion
using the {\bf k}-space formulation of Laks {\em et al.}.
\cite{Laks92}
Rather than a cluster expansion of $\Delta H_{\sigma}$, we will
expand with respect to a reference energy:
\begin{equation}
\label{e.ref}
\Delta E_{\rm CE}(\sigma) = \Delta H^{\rm LDA}(\sigma) - E_{\rm ref}
\end{equation}
We will separate the CE into two parts:
(i) the terms corresponding to pair
interactions with arbitrary separation
will be conveniently summed using the reciprocal-space
concentration-wave formalism, and (ii) all terms but the
pairs will be cast in real-space:
\begin{equation}
\label{ce.mixed}
\Delta E_{\rm CE}(\sigma) = \sum_{\bf k} J({\bf k}) |S({\bf k},\sigma)|^2
+ \sum_{f} D_f \hspace{3pt} J_f \hspace{3pt}
\overline{\Pi}_f(\sigma).
\end{equation}
The first summation includes all pair figures
and the second summation includes only non-pair figures.
In the reciprocal-space summation in Eq.\ (\ref{ce.mixed}),
$J({\bf k})$ and $S({\bf k},\sigma)$ are the lattice Fourier
transforms of the real-space pair interactions and
spin-occupation variables, $J_{ij}$ and $\hat{S}_i$,
respectively, and the spin-occupation variables take the
value $\hat{S}_i=-1(+1)$ is the atom at site $i$ is Ni(Au).
The function $J({\bf k})$ is required to be a smooth
function by minimizing the integral of the gradient of
$J({\bf k})$.
The real-space summation of Eq.\ (\ref{ce.mixed}) is over
$f$, the symmetry-distinct non-pair figures
(points, triplets, etc.), $D_f$ is the number of figures
per lattice site, $J_f$ is the Ising-like interaction for the figure
$f$, and $\overline{\Pi}_f$
is a product of the variables $\hat{S}_i$
over all sites of the figure $f$,
averaged over all symmetry equivalent figures of lattice sites.

The reference energy of Eq.\ (\ref{e.ref})
is chosen to contain infinite-range real-space
elastic interaction terms.
Subtracting these long-range terms from $\Delta H^{\rm LDA}_{\sigma}$
before cluster expanding removes the ${\bf k}\rightarrow0$
singularity, and thus significantly enhances the convergence
of the CE. \cite{Laks92}  The form used for $E_{\rm ref}$ is
\begin{equation}
\label{e.ref.2}
E_{\rm ref} = \frac{1}{4x(1-x)} \sum_{{\bf k}}
\Delta E^{\rm eq}_{\rm CS}(\hat{k},x) |S({\bf k},\sigma)|^2
\end{equation}
where $\Delta E^{\rm eq}_{\rm CS}(\hat{k},x)$ is the
equilibrium {\em constituent strain energy},
defined as the energy change when the bulk solids Ni and Au
are deformed from their equilibrium cubic lattice constants
$a_{Ni}$ and $a_{Au}$ to a common lattice constant
$a_{\perp}$ in the direction perpendicular to $\hat{k}$.
$\Delta E^{\rm eq}_{\rm CS}(\hat{k},x)$ can thus be
written as the minimum of the following
expression with respect to $a_{\perp}$:
\begin{equation}
\label{cs.q}
\Delta E_{\rm CS}^{\rm eq}(\hat{k},x) = (1-x) q^{\rm Ni}(a_{\perp},\hat{k}) 
\Delta E^{\rm Ni}_{\rm bulk}(a_{\perp})
+ x q^{\rm Au}(a_{\perp},\hat{k}) 
\Delta E^{\rm Au}_{\rm bulk}(a_{\perp}).
\end{equation}
where $q^{(\lambda)}(a_{\perp},\hat{k})$ is given by Eq. (\ref{cont.elasticity}).

The final expression used for the formation energy of any
configuration $\sigma$ is then
\begin{eqnarray}
\label{eform.ce}
\Delta H(\sigma) &=& \sum_{\bf k} J({\bf k}) |S({\bf k},\sigma)|^2
+ \sum_{f} D_f \hspace{3pt} J_f \hspace{3pt} 
\overline{\Pi}_f(\sigma) \nonumber \\
&& + \frac{1}{4x(1-x)} \sum_{{\bf k}}
\Delta E^{\rm eq}_{\rm CS}(\hat{k},x) |S({\bf k},\sigma)|^2
\end{eqnarray}
The following {\em input} is needed to construct this Hamiltonian for Ni-Au: 
(i) the formation energies of a set of ordered fcc-based 
compounds (required to fit the values of $J({\bf k})$ and $J_f$), and 
(ii) the epitaxial energies of fcc Ni and fcc Au (required to
compute the anharmonic values of $\Delta E_{\rm CS}^{\rm eq}(\hat{k},x)$).
The {\em output} is a Hamiltonian [Eq.\ (\ref{eform.ce})] which  
(i) predicts the energy of any fcc-based configuration (i.e., not
only ordered compounds) even 1000-atom cells or larger,
(ii) possesses the accuracy of fully-relaxed, 
full-potential LDA energetics, and
(iii) is sufficiently simple to evaluate that it can be
used in Monte Carlo simulations, and thereby extend
LDA accuracy to finite temperatures.

\subsection{Anharmonic calculation of constituent strain}
\label{anharm.cs}

Laks {\em et al.} \cite{Laks92} demonstrated that the
calculation of $\Delta E^{\rm eq}_{\rm CS}(\hat{k},x)$ of Eq. (\ref{cs.q})
is significantly
simplified if one uses harmonic continuum elasticity theory 
[i.e., insert Eqs.\ (\ref{q.harm})-(\ref{gamma}) 
into Eq.\ (\ref{cs.q})];
However, we have already seen evidence of anharmonic
elastic effects in Ni-Au.  Thus, we have performed
LDA calculations of 
$q(a_{\perp},\hat{k})$ directly from its definition in
Eq.\ (\ref{cont.elasticity}), rather than using the harmonic
approximation in Eq.\ (\ref{gamma}).
In Fig.\ \ref{niau.q}, we show the results of the 
LAPW calculations of 
$q^{\rm Ni}(a_{\perp},\hat{k})$ and $q^{\rm Au}(a_{\perp},\hat{k})$
for six principle directions:
(100), (111), (110), (201), (311), and (221).
It is clear that the calculated values of $q$ are
not independent of $a_{\perp}$, but rather show a marked and non-trivial
dependence on the perpendicular lattice constant.  Thus, 
the lattice mismatch in Ni-Au appears to be too large for
a {\em harmonic} continuum model of elasticity to be accurate.
In particular,
the value of $q^{\rm Ni}(a_{\perp},100)$ is quite low upon 
expansion, indicating that Ni is elastically extremely soft in
this direction.  Au, on the other hand, becomes softest in the
(201) direction for significant compression.  
In a separate publication, \cite{Ozolins97} we will demonstrate
that the anharmonic effects can be cast analytically in terms
of the harmonic expressions of Eqs.\ (\ref{q.harm})-(\ref{gamma})
by extending the expansion of $\gamma(\hat{k})$: 
\begin{equation}
\gamma(a_{\perp},\hat{k}) = \sum_L a_L(a_{\perp}) K_L(\hat{k})
\end{equation}
to include angular momenta $L$=6,8, and 10 
with the coefficients $a_L(a_{\perp})$ obtained from LDA calculations
rather than the $L$=0,4 expression of Eq. (\ref{gamma})
used before. \cite{Laks92} 

The results for
$q^{\rm Ni}(a_{\perp},\hat{k})$ and $q^{\rm Au}(a_{\perp},\hat{k})$
are used to numerically minimize Eq.\ (\ref{cs.q}) and hence to
find $\Delta E_{\rm CS}^{\rm eq}(\hat{k},x)$.  
The results for the CS energies
are shown in Fig.\ \ref{niau.cs}.  Here, also, the anharmonic
effects are seen quite strongly as $\Delta E_{\rm CS}^{\rm eq}(\hat{k},x)$
for some directions cross with other directions
and asymmetries of the various directions are not all the same
(effects which could not occur in the harmonic model).
The most prominent feature of $\Delta E_{\rm CS}^{\rm eq}(\hat{k},x)$ is that
(100) is the softest elastic direction, which stems from the elastic
softness of Ni along this direction.  Ni being soft and Au being relatively
hard along (100) leads to Ni(Au) being highly distorted (nearly undistorted)
for long-period (100) Ni-Au superlattices, and also leads to the
marked asymmetry in $\Delta E_{\rm CS}^{\rm eq}(100,x)$ towards the
Ni-rich compositions.  Similar arguments can be applied to explain
the opposite asymmetry of the (201) strain.

For $E_{\rm ref}$ to be useful in the {\bf k}-space CE, one must be
able to know this energy for {\em all directions}, not merely the ones
for which it was calculated.
To obtain such a useful form, we fit
the constituent strain results of Fig.\ \ref{niau.cs} to a series of Kubic
harmonics (0-10th order) consistent with cubic symmetry ($L$ = 0,4,6,8,10).  
This procedure
provides not only a good fit of the calculated strain data, but also an
analytic form to obtain the values of $\Delta E_{\rm CS}^{\rm eq}(\hat{k},x)$
for all directions.

\subsection{Stability of the cluster expansion}

Using the calculated formation energies $\{\Delta H_{\sigma}\}$ 
(Table \ref{niau.eform}) and
the anharmonic CS strain energy (Fig.\ \ref{niau.cs}), 
we then fit the coefficients $J({\bf k})$ and $\{J_f\}$
of the {\bf k}-space CE using Eq.\ (\ref{ce.mixed}).
We used all 33 calculated structures in the fit of the expansion,
which included 20 pair, 5 triplet, and 3 quadruplet interactions.
The standard deviation of the fitted energies relative to their
LAPW values is 5.3 meV/atom, which is the same order of magnitude
as the numerical uncertainties in LAPW.
The results for pair and multibody interactions are shown in Fig.\ \ref{niau.jf}.

In order for the expansion to have a useful predictive capability,
tests must be performed to assess the stability of the fit:

{\em Changing the number of interactions}:  We performed tests of the
stability of the fit with respect to the number of pair interactions,
$N_{\rm pairs} = (1-50)$.  Figure \ref{niau.error}
shows the standard deviation of the fit as a function of the number
of pairs interactions included.  It is clear that the fit is well converged
for $N_{\rm pairs}=20$.  We also tested the stability of the fit with respect
to inclusion of more multibody interactions
than are shown in Fig.\ \ref{niau.jf}:  Including three additional triplet
figures in the fit resulted in no change of the standard deviation of the
fit, the added interactions had values $<2$ meV/atom, and the original
interactions were changed by less than 2 meV/atom.  Thus, the fit
is stable with respect to the figures included (both pair and multibody).

{\em Changing the number of structures}:  We also performed tests of the
predictive ability of the fit by removing some structures from the fit.
First, we removed three structures which were originally 
fit quite well: $Z2$, $\beta 2$, and $L1_2$ (NiAu$_3$).  Removing these
structures from the input set resulted in their energies changing
by $\lesssim$1 meV/atom.  However, a much more critical test of the
fit is to remove the structures which are fit most poorly:  
$SQS14_a$ and $SQS14_b$.  Removing these structures from the fit
changes their energies by only $\sim$2-3 meV/atom.  Thus, we are
confident that the present {\bf k}-space CE fit is both stable
and predictive.

\section{Results of Current Calculations}

\subsection{Mixing enthalpy:  How good are previous calculations?}

Using the {\bf k}-space cluster expansion in combination with
a mixed real/reciprocal space Monte Carlo code (canonical), 
one can obtain
thermodynamic properties of Ni-Au alloys.  Figure \ref{niau.mc}
shows the mixing enthalpy as a function of temperature,
$\Delta H_{\rm mix}(T)$.  Monte Carlo calculations were
performed for a 16$^3$=4096 atom cell, with 100 Monte Carlo
steps per site for averages.  The simulation was started
at an extremely high temperature, and slowly cooled down
using a simulated annealing algorithm.
Also shown in Fig.\ \ref{niau.mc} is the value of the
mixing energy of the completely random alloy.  The difference
between the Monte Carlo calculated $\Delta H_{\rm mix}(T)$
and the random alloy energy is precisely the energetic effect
of short-range order.  We have fit the values of
$\Delta H_{\rm mix}(T)$ to linear and quadratic functions
of $\beta=1/k_BT$ to extrapolate the values down in temperature
below the point at which coherent phase separation occurs 
in the simulation.  (Both fits gave virtually identical 
results, so the linear fit is used here and below.)
This allows us to ascertain the value
of the mixing enthalpy at 1100 K, near the temperature where
this quantity has been experimentally measured.
These results are tabulated in Table \ref{niau.mix}, which
shows both the 
effects of atomic relaxation ($\sim 100$ meV/atom) and
SRO ($\sim 25$ meV/atom) on the mixing
enthalpy, and compares the value of atomically relaxed
and short-range ordered mixing energy with those values
from experiment.  One can see that by taking into account
{\em both} relaxation and SRO, LDA produces
a value for the mixing energy which is only different
from experiment by 15-20 meV/atom.  Thus, we conclude from
this comparison that high quality LDA calculations provide
accurate energetics for the Ni-Au system.  

The preceding discussion leads to a number of conclusions
regarding previous calculations of $\Delta H_{\rm mix}$:

(i) Since relaxation reduces $\Delta H_{\rm mix}$ by
$\sim$100 meV/atom, the unrelaxed $\Delta H_{\rm mix}$ values
(``d'' in Table \ref{comp.niau}) have to be reduced by this
amount to appropriately compare with experiment.

(ii) Since SRO reduces $\Delta H_{\rm mix}$ by
$\sim$25 meV/atom, the results of previous calculations
that omitted SRO (all except ``i'' in Table \ref{comp.niau})
have to be adjusted accordingly.

(iii) In light of the fact that the empirical potential-based 
and ASA-based methods (LMTO and ASW)
were shown to be inaccurate with respect to full-potential
LDA methods for {\em unrelaxed, ordered compounds}
(Table \ref{comp.niau.ord}), the results of {\em relaxed, mixing energies
of random alloys} appear to be questionable using these
schemes.  

\subsection{Configurational or non-configurational entropy?}

From the fit of the Monte Carlo data in Fig.\ \ref{niau.mc}, one can
find the configurational entropy of the Ni$_{0.5}$Au$_{0.5}$
disordered phase by
integrating the energy down from infinite temperature (where
the configurational entropy is known):
\begin{equation}
\Delta S(T) = \Delta S(T=\infty) + E(T)/T - k_B\int_0^{\beta} E(\beta)d\beta
\end{equation}
The configurational entropy obtained from thermodynamic integration
in this way is 
\begin{equation}
\Delta S_{\rm conf.}({\rm Ni}_{0.5}{\rm Au}_{0.5},T=1100 K)=0.56 k_B,
\end{equation}
compared to the ``ideal'' (infinite temperature) value of 
\begin{equation}
\Delta S_{\rm conf.}({\rm Ni}_{0.5}{\rm Au}_{0.5},T\rightarrow\infty)=0.69 k_B, 
\end{equation}
This calculated value for the {\em configurational} entropy
of mixing can be compared with the experimentally measured
values of {\em total} entropy of mixing:
Calorimetric measurements give $\Delta S(T=1150 K)=1.04k_B$ \cite{Hultgren} while
EMF measurements give $\Delta S(T=1173 K)=1.08k_B$ \cite{emf}.
Thus, we can obtain an estimate of the {\em non-configurational}
entropy, and find it to be large:
$\Delta S_{\rm non-conf.}(T \sim 1100 K) = 1.04-0.56 = 0.48k_B$.
This non-configurational entropy is hence 
responsible for $T_{\rm MG}$ being so small experimentally,
compared to all the theoretical results.  In fact, if we
use the calculated $\Delta H_{\rm mix}$ = 93 meV/atom and
the combined ``experimental/calculated'' 
$\Delta S_{\rm non-conf.}$ = 0.48$k_B$ in the following
formula:
\begin{equation}
T_{\rm MG} = \frac{2\Delta H_{\rm mix}}{k_B+2\Delta S_{\rm non-conf.}}
\end{equation}
we obtain $T_{\rm MG} \sim 1100$ K and $k_BT_{\rm MG}/\Delta H_{\rm mix}=1.02$, 
much closer to the experimental values
($T_{\rm MG} \sim 1083$ K and $k_BT_{\rm MG}/\Delta H_{\rm mix}=1.2$) 
than using the above formula neglecting
non-configurational entropy 
($T_{\rm MG} \sim 2150$ K and $k_BT_{\rm MG}/\Delta H_{\rm mix}=2.0$).  

From this consideration of non-configurational effects, one
should conclude that the accuracy of a calculation
with configurational degrees of freedom only (as is done in most
of the previous calculations \cite{note.2}), should
be determined by looking at the {\em energetics}, not the {\em transition
temperatures}.  Thus, previous calculations which give ``good''
transition temperatures do so precisely because they have
``bad'' energetics.  

\subsection{Short-range order of Ni$_{1-x}$Au$_x$ solid solutions}

Using the {\bf k}-space CE and Monte Carlo, we may also compute
the SRO of disordered Ni$_{1-x}$Au$_x$ alloys.
We show the results of our SRO simulations for
Ni$_{0.4}$Au$_{0.6}$ in
Fig.\ \ref{niau.sro}.  For the SRO Monte Carlo
calculations, a cell of 24$^3$=13824 atoms was
used, with 100 Monte Carlo steps for equilibration,
with averages taken over the subsequent 500 steps.
Several calculations and measurements
of the SRO exist in the literature:  Wu and
Cohen \cite{Cohen2} used diffuse x-ray scattering to
deduce the atomic SRO of Ni$_{0.4}$Au$_{0.6}$ at
$T$=1023 K.
The measured diffuse intensity due to SRO
must be separated from all the other contributions which give
rise to diffuse intensity, and for this purpose,
Wu and Cohen used 25 real-space Fourier shells of SRO parameters,
and found the rather surprising result that
the peak intensity in reciprocal space due to SRO is of
ordering-type and occurs at the point
${\bf k}_{\rm SRO}$=(0.6,0,0), rather than
${\bf k}_{\rm SRO}$=(0,0,0) which would be expected for a clustering
alloy.  Several authors have tried to account for this ordering 
nature of the SRO: Lu and Zunger \cite{Lu94}
calculated the SRO (using 21 real-space shells) 
and found peaks
at $\sim$(0.8,0,0) whereas Asta and Foiles \cite{Asta}
used an embedded atom method and found the SRO
(using 8 real-space shells) to peak at $\sim$(0.5,0,0). 
Our calculations for the SRO of
Ni$_{0.4}$Au$_{0.6}$ are given in Fig.\ \ref{niau.sro}.
We have calculated the SRO at $T$=2300 K, above
the miscibility gap temperature for our alloy Hamiltonian.
We find that, using 8, 25, and 100 shells, the SRO peaks at
(0.65,0,0), (0.40,0,0), and (0.38,0,0) respectively, in good agreement
with both the measurements of Wu and Cohen [${\bf k}_{\rm SRO}$=(0.6,0,0)
for 25 shells] and also with previous calculations.

Equation (\ref{eform.ce}) shows that the alloy Hamiltonian used
in the Monte Carlo simulations is composed of three parts:
the pair interaction terms, the multibody interaction terms,
and the constituent strain terms.
It is interesting to see the effect of each of these
portions of the alloy Hamiltonian on SRO.
Thus, in addition to the ``full'' calculations,
which contain pairs, multibodies, and constituent strain in the
alloy Hamiltonian, we have also computed the SRO with (i) the CS energy
only, and (ii) the CS energy plus the pair interactions.
These results are shown in Fig.\ \ref{niau.sro.cs}.
(Because the CS energy is non-analytic in reciprocal space about
the origin, many Fourier coefficients are required to 
converge the SRO of CS alone, thus we show only results using
100 shells of parameters in Fig.\ \ref{niau.sro.cs}.)
One can see that the SRO with CS only
is dominated by almost constant streaks of intensity along
the $\Gamma-X$ line, and very little intensity elsewhere.
This SRO pattern is understandable when one considers that the
constituent strain at this composition (Fig.\ \ref{niau.cs})
is much softer (much lower in energy) 
in the (100) direction than along any other
direction.  Thus, (100)-type fluctuations in the random alloy are 
be energetically favored, and because the constituent strain is
dependent only on direction and not on the length of the wavevector,
one should expect that all fluctuations along the (100) will
occur roughly equally, regardless of the length of the wavevector.
This is precisely what we see in Fig.\ \ref{niau.sro.cs}.
Contrasting this SRO using CS only with that
calculated both CS energy and pair interactions
(but not multibody interactions) shows that the pair interactions
create a peak in intensity along the $\Gamma-X$ line, but
significantly closer to $\Gamma$ than the peak intensity
using the ``full'' alloy Hamiltonian.  Thus, while the
effect of pairs is to create a peak near the $\Gamma$ point,
the multibody interactions move this peak out from $\Gamma$
towards the $X$-point.

\subsection{Standard inverse Monte Carlo would give unphysical
interaction energies:  a challenge}

The statistical problem we have solved here involves the calculation
of the alloy SRO at high temperature for given alloy Hamiltonian 
($\{J_{ij}\}$, $\{J_f\}$, and $\Delta E_{\rm CS}$).
However, 
a popular technique used to study phase stability in alloys
involves the ``inverse'' problem of determining a set of 
pair-only interactions $\{\tilde{J}_{ij}\}$ from a measured or
calculated SRO pattern, and the subsequent use of these
pair interactions to determine thermodynamic properties
other than the SRO.  In fact, $\{\tilde{J}_{ij}\}$ are
often used to determine $\Delta H_{\rm mix}$ or phase
stability.  As we have mentioned in the introduction
and described more fully in Ref. \onlinecite{Wolverton97},
inverting the SRO always removes information on energy
terms that are SRO-independent, e.g., the volume
deformation energy $G(x)$.  This loss
prevents, in principle, the interactions deduced from SRO from
being applied to predict physical properties which depend on $G(x)$,
such as $\Delta H_{\rm mix}$.
For example, in the case of Ni-Au, the SRO is of ordering-type.
Thus, we expect that inverting the SRO of Ni-Au (e.g., via inverse
Monte Carlo) would produce interactions $\{\tilde{J}_{ij}\}$ which
are of ordering-type, and using these interactions to predict
the mixing enthalpy would result in the unphysical result 
$\Delta H_{\rm mix}<0$.

One might suspect that by changing the temperature, 
one could obtain a shift of the SRO from ordering- 
to clustering-type, and thus, the inverse technique would then 
produce interactions
which would correctly give $\Delta H_{\rm mix}>0$.
However, we have computed the SRO for several temperatures,
and find no evidence of a shift in SRO to clustering-type.

A test of our expectations 
by any of the practitioners of inverse Monte Carlo
would certainly be welcomed.
To that end, our SRO calculations are available for use as
input to inverse Monte Carlo to extract interactions.  
These SRO calculations are available for a variety of 
compositions and temperatures, each with a large number
of real-space SRO parameters.
It would be of great interest to see whether the
interactions extracted from inverting the SRO of Ni-Au
would produce the correct sign of $\Delta H_{\rm mix}$.

\vspace{20pt}

\begin{center}
{\bf Acknowledgements}
\end{center}

This work was supported by the Office of Energy Research
(OER) [Division of Materials Science of the Office of Basic Energy
Sciences (BES)], U. S.  Department of Energy, under contract No.
DE-AC36-83CH10093.  The authors would like to thank Dr. M. Asta
for providing the EAM values in Table \ref{comp.niau.ord}
and Mr. D. Morgan for communicating his results to us prior
to publication.

%
%

%
%
%
%

\begin{figure}[tb]
\caption{LAPW calculations of $q^{(\lambda)}(a_{\perp},\hat{k})$ of 
Eq.\ (\protect\ref{cont.elasticity})
for Ni-Au.  Shown are (a) $q^{\rm Ni}$ and
(b) $q^{\rm Au}$ for six principle
directions.}
\label{niau.q}
\end{figure}

\begin{figure}[tb]
\caption{LAPW calculations of $\Delta E_{\rm CS}(\hat{k},x)$ for Ni-Au  
for six principle directions.}
\label{niau.cs}
\end{figure}

\begin{figure}[tb]
\caption{(a) Pair and (b) multibody interaction energies for Ni-Au.
The multibody figures are defined by the following lattice
sites, in units of $a$=2 (the origin is contained in all figures):
$J_3$ - (110),(101),
$K_3$ - (110),(200),
$N_3$ - (200),(002),
$P_3$ - (110),(103),
$Q_3$ - (110),(220),
$J_4$ - (110),(101),(011),
$K_4$ - (110),(101),(200), and
$L_4$ - (110),(101),(211).}
\label{niau.jf}
\end{figure}

\begin{figure}[tb]
\caption{Cluster expansion fitting error in Ni-Au versus the number of
pair interactions included in the fit.}
\label{niau.error}
\end{figure}

\begin{figure}[tb]
\caption{$\Delta H(T)$ computed for Ni$_{0.5}$Au$_{0.5}$ from a combination
of the {\bf k}-space cluster expansion and Monte Carlo simulations.}
\label{niau.mc}
\end{figure}

\begin{figure}[tb]
\caption{Monte Carlo-calculated 
short-range order of Ni$_{0.4}$Au$_{0.6}$ in the ($hk0$) plane
using (a) 8, (b) 25, and (c) 100 shells of Warren Cowley SRO parameters.
Peak intensity is red shaded contour while the lowest contours are
shaded blue.  Contours are separated by 0.1 Laue unit in each plot.}
\label{niau.sro}
\end{figure}

\begin{figure}[tb]
\caption{Monte Carlo-calculated 
short-range order of Ni$_{0.4}$Au$_{0.6}$ using
(a) constituent strain terms only,
(b) constituent strain and pair terms, and
(c) constituent strain, pair, and multibody terms
in the alloy Hamiltonian.
Peak intensity is red shaded contour while the lowest contours are
shaded blue.  Contours are separated by 0.1 Laue unit in each plot.}
\label{niau.sro.cs}
\end{figure}

%
%
%
%

{\squeezetable
\begin{table}
\caption{Summary of energy calculations performed for Ni$_{1-x}$Au$_x$ 
alloys.  Shown are the methods used to compute
$T$=0 energetics, 
as well as the type of cluster expansion (CE) and statistics used.
Also given is the mixing energy of the $T \rightarrow \infty$
random alloy near $x$=1/2, and the calculated miscibility gap
temperature, if available.  FLAPW = full-potential linearized
augmented plane wave method, FLMTO = full-potential linear
muffin-tin orbital method, ASW = augmented spherical wave method,
LMTO-ASA = linear muffin-tin orbital method in the atomic sphere
approximation, EAM = embedded atom method, MC = Monte Carlo, CVM = 
cluster variation method, MF = mean-field, SOE = second-order
expansion.}
\label{comp.niau}
\begin{tabular}{l|ddd|dd}
&&{\small Method}&&
\multicolumn{2}{c}{\small Results}\\
&&Cluster&&\\
Authors&T=0 Energy&Expansion&Statistics&
$\Delta H_{\rm mix}^{\rm fcc}$& $T_{\rm MG}$(K) \\
&&Technique&&\\
\tableline
Wolverton and Zunger \tablenotemark[1]& FLAPW & {\bf k}-space CE & MC & +118 &
        \\
Lu and Zunger \tablenotemark[2]& FLAPW & $\epsilon-G$ & MC & +127     \\
Deutsch and Pasturel \tablenotemark[3] & FLMTO & $\epsilon-G$ & none & +136     \\
Takizawa, Terakura, and Mohri\tablenotemark[4]& ASW  & CW & CVM & +170 &          \\
Amador and Bozzolo\tablenotemark[5] & LMTO-ASA & CW & CVM & +150     \\
Colinet {\em et al.}\tablenotemark[6] & LMTO-ASA & $\epsilon-G$ & CVM  & +67&1200-1400  \\
Morgan and de Fontaine \tablenotemark[7] & LMTO-ASA + & $\epsilon-G$ 
& CVM &  +98 & 2330\\
&``Elastic Springs''&&&\\
Eymery {\em et al.}\tablenotemark[8] & Empir. Potential & Simulation & none &  +60    &\\
Tetot and Finel\tablenotemark[9] & Empir. Potential & Simulation & MC & +48\tablenotemark[13]&950\\
Deutsch and Pasturel \tablenotemark[3] & Empir. Potential & Simulation & none &  +83      \\
Asta and Foiles\tablenotemark[10]& EAM & SOE & MC/MF & +78&2460\\
\tableline
Expt. (Calorimetry) $T$=1150 K \tablenotemark[11]&    & &   & +76      \\
Expt. (EMF) $T$=1173K \tablenotemark[12]        &   & &   & +77      \\
Expt. (Phase Diagram)                           &   & &   &    &1083\\
\end{tabular}
\tablenotetext[1]{Present results.}
\tablenotetext[2]{Ref. \cite{Lu94}}
\tablenotetext[3]{Ref. \cite{Pasturel}}
\tablenotetext[4]{Ref. \cite{Mohri}}
\tablenotetext[5]{Ref. \cite{Amador}}
\tablenotetext[6]{Ref. \cite{Colinet}}
\tablenotetext[7]{Ref. \cite{Morgan}}
\tablenotetext[8]{Ref. \cite{Eymery}}
\tablenotetext[9]{Ref. \cite{Finel}}
\tablenotetext[10]{Ref. \cite{Asta}}
\tablenotetext[11]{Ref. \cite{Hultgren}}
\tablenotetext[12]{Ref. \cite{emf}}
\tablenotetext[13]{at $T$=1150 K}
\end{table}
}

\eject

{\squeezetable
\widetext
\begin{table}
\caption{Comparison of formation enthalpies $\Delta H_f(\sigma)$
for Ni-Au ordered compounds. Nomenclature for the ordered structures
is the same as that used in Ref. \protect\onlinecite{Zunger94}.
All energies in meV/atom.
Numbers in parentheses indicate unrelaxed energies.}
\label{comp.niau.ord}
\begin{tabular}{cccccccc}
&Fully Relaxed&Partially Relaxed&&&&Empirical\\
Structure&
FLAPW  \tablenotemark[1]&
FLAPW  \tablenotemark[2]&
FLMTO \tablenotemark[3]&
ASW   \tablenotemark[4]&
LMTO  \tablenotemark[5]&
Potential  \tablenotemark[3] &
EAM  \tablenotemark[10] \\
\tableline
NiAu ($L1_0$)      & +76.1 (+98.1)   & +76.8  & +79.4 (+96.4) 
		   &(+59)            &(+116.6)&$+$57.9($+$73.9) &$+$21.4($+$91.1)\\
Ni$_2$Au$_2$ ($Z2$)& +70.2 (+286.7)  & +124.3 &+123.1 (+300.1)
		   &                 &(+213.4)&$+$62.3($+$127.7)&$-$130.3($+$208.6)\\
NiAu ($L1_1$)      & +166.8 (+192.3) & +167.6 &+175.4    
		   &                 &(+177.9)&                 &$+$72.9($+$159.7)\\
NiAu ($``40''$)    & +84.8 (+93.5)   & +83.8  & +89.9    
		   &                 &(+114.3)&                 &$-$1.9($+$96.4)\\
Ni$_3$Au ($L1_2$)  & +77.5           & +75.5  & +80.7    
		   & +42             &+92.4   &$+$58.1&$+$77.1\\
Ni$_3$Au ($D0_{22}$)&+75.0 (+75.0)   &        &  +81.5   
		   &                 &(+95.3)\\
NiAu$_3$ ($L1_2$)  &+78.9            & +78.2  &  +78.0   
		   & +52             &+89.4   &$+$54.1&$+$86.1\\
NiAu$_3$ ($D0_{22}$)&+68.6 (+68.7)   &        &  +68.0   
		   &                 &(+76.4)\\
\end{tabular}
\tablenotetext[1]{Present results. 
Complete atomic relaxation via quantum mechanical forces and
total-energy minimization.}
\tablenotetext[2]{Ref. \cite{Lu94}.
     Partial atomic relaxation via continuum elasticity, using 
     Eqs.\ (\protect\ref{cont.elasticity})-(\protect\ref{gamma}).}
\tablenotetext[3]{Ref. \cite{Pasturel}}
\tablenotetext[4]{Ref. \cite{Mohri}}
\tablenotetext[5]{Ref. \cite{Morgan}.
LMTO-ASA with sphere radii chosen to minimize charge transfer.}
\tablenotetext[10]{Ref. \cite{Asta}}
\end{table}

}

{\squeezetable
\widetext
\begin{table}
\caption{Listing of the LAPW calculated unrelaxed and relaxed 
$\Delta H(\sigma)$ 
[in meV/atom] for Ni$_{1-x}$Au$_x$. 
Many of the structures
calculated here can be characterized as a (Ni)$_p$(Au)$_q$
superlattice of orientation $\hat{k}$.  
We use the nomenclature of Ref. \protect\onlinecite{Zunger94} for
structure names.}
\label{niau.eform}
\begin{tabular}{cddddd}
Orientation   &  [001]  & [011]   & [012]   & [111] & [113] \\
formula       &         &         &         &       &       \\
\tableline
 $AB$     &   $L1_0$ &   $L1_0$  &   $L1_0$  &   $L1_1$  & $L1_1$    \\
Unrelaxed & $+$98.1  & $+$98.1   & $+$98.1   & $+$192.3  & $+$192.3  \\
Relaxed   & $+$76.1  & $+$76.1   & $+$76.1   & $+$166.8  & $+$166.8  \\
CE(Relaxed)&$+$74.8  & $+$74.8   & $+$74.8   & $+$167.1  & $+$167.1  \\
\tableline
 $A_2B$   & $\beta$1 & $\gamma$1 &           & $\alpha$1&    \\
Unrelaxed & $+$207.8 & $+$123.3  &           & $+$288.5  &   \\
Relaxed   & $+$105.7 & $+$98.9   &           & $+$202.2  &   \\
CE(Relaxed)&$+$105.9 & $+$102.4  &           & $+$208.4  &   \\
\tableline
 $AB_2$   & $\beta$2 & $\gamma$2 &           & $\alpha$2&   \\
Unrelaxed & $+$151.7 & $+$126.3  &           & $+$200.9  &   \\
Relaxed   & $+$38.3  & $+$102.6  &           & $+$100.9  &   \\
CE(Relaxed)&$+$37.8  & $+$98.8   &           & $+$94.5   &   \\
\tableline
 $A_3B$   &    Z1    & Y1        & $DO_{22}$ & V1        & W1        \\
Unrelaxed & $+$221.7 & $+$148.5  &  $+$75.0  & $+$290.8  &           \\
Relaxed   & $+$89.9  & $+$99.2   &  $+$75.0  & $+$193.7  & $+$125.7  \\
CE(Relaxed)&$+$94.3  & $+$91.3   &  $+$69.1  & $+$189.6  & $+$120.8  \\
\tableline
 $AB_3$   &    Z3    & Y3        & $DO_{22}$ & V3        & W3        \\
Unrelaxed & $+$142.0 & $+$104.1  & $+$68.7   & $+$172.8  &           \\
Relaxed   & $+$32.4  & $+$78.7   & $+$68.6   & $+$83.0   & $+$88.4   \\
CE(Relaxed)&$+$28.2  & $+$77.7   & $+$67.6   & $+$79.1   & $+$83.2   \\
\tableline
 $A_2B_2$ &    Z2    &    Y2     &  ``40''   &    V2     &  W2       \\
Unrelaxed & $+$286.7 & $+$192.3  &   $+$93.5 & $+$335.8  & $+$144.2  \\
Relaxed   & $+$70.2  & $+$96.6   &   $+$84.8 & $+$162.4  & $+$93.6   \\
CE(Relaxed)&$+$69.9  & $+$101.1  &   $+$88.3 & $+$166.7  & $+$99.3   \\
\tableline
$A_pB_p (p \rightarrow \infty)$ \\
Unrelaxed & $+$576.2 & $+$576.2  &   $+$576.2& $+$576.2   & $+$576.2  \\
Relaxed   & $+$30.8  & $+$117.7  &   $+$84.8 & $+$173.8   & $+$119.8   \\
CE(Relaxed)&$+$30.8  & $+$116.1  &   $+$86.8 & $+$172.5   & $+$117.9   \\
\tableline
\multicolumn{4}{c}{Other Structures} \\
          & $L1_2$ ($A_3B$)  & $L1_2$ ($AB_3$)  & $D7$ ($A_7B$)  & $D7_b$ 
	  ($A_7B$)\\
Unrelaxed & $+$77.5   & $+$78.9    & $+$82.9   & 56.8 & \\
Relaxed   & $+$77.5   & $+$78.9    & $+$82.9   & 56.8 & \\
CE(Relaxed)&$+$80.7   & $+$78.6    & $+$98.5   & 57.6 & \\
\tableline
          & SQS14$_a$ ($A_6B_2$)  & SQS14$_b$ ($A_2B_6$)  & Z6 ($A_3B_3$ - 100) & Z5 ($A_2B_3$ - 100)\\
Unrelaxed & $+$183.2   & $+$118.2    & $+$355.5  & $+$273.3 \\
Relaxed   & $+$96.8    & $+$59.8    &  $+$63.2   & $+$57.1  \\
CE(Relaxed)&$+$81.5    & $+$75.0    &  $+$62.5   & $+$57.9  \\
\tableline
\end{tabular}
\end{table}
}

\eject

%
%

\widetext
\begin{table}
\caption{$\Delta H_{\rm mix}$
for Ni$_{0.5}$Au$_{0.5}$. All energies in meV/atom.
SQS-4 refers to a 4-atom special quasi-random structure (Y2).
This table shows the effects of relaxation (first line minus second
line) and short-range order (third line minus fourth line) on
the mixing energy.}
\label{niau.mix}
\begin{tabular}{cc}
 & $\Delta H_{\rm mix}$ \\
\tableline
SQS-4 Unrelaxed($T = \infty$)         &  $+$192 \\
SQS-4 Relaxed($T = \infty$)           &  $+$97  \\
\tableline
CE    Relaxed($T = \infty$)           &  $+$118 \\
CE    Relaxed($T=1100 K$)            &  $+$93    \\
\tableline
Expt. (Calorimetry) $T$=1150 K & $+$76 \\
Expt. (EMF) $T$=1175 K         & $+$77 \\
\end{tabular}
\end{table}


\begin{references}

\bibitem{Hultgren}	
			{\em Selected Values of Thermodynamic Properties of
		     	Metals and Alloys}, R. R. Hultgren {\em et al.} eds. 
     			(Wiley, New York, 1963). 

\bibitem{Cohen2}	T. B. Wu and J. B. Cohen,
			Acta Metall. {\bf 31}, 1929 (1983).

\bibitem{Moss1}		B. Golding and S. C. Moss,
			Acta Metall. {\bf 15}, 1239 (1967).

\bibitem{Moss2}		B. Golding, S. C. Moss, and B. L. Averbach,
			Phys.\ Rev.\ {\bf 158}, 637 (1967).

\bibitem{Cohen1}	T. B. Wu, J. B. Cohen, and W. Yelon,
			Acta Metall. {\bf 30}, 2065 (1982).

\bibitem{Cohen3}	T. B. Wu and J. B. Cohen,
			Acta Metall. {\bf 32}, 861 (1984).

\bibitem{Cook}		H. E. Cook and D. de Fontaine,
			Acta Metall. {\bf 17}, 915 (1969).

\bibitem{Lu94}          Z. -W. Lu and A. Zunger, Phys.\ Rev.\ B
			{\bf 50}, 6626 (1994);

\bibitem{e-G}		L. G. Ferreira, A. A. Mbaye, and A. Zunger,
			Phys.\ Rev.\ B {\bf 37}, 10547 (1988).

\bibitem{imc}		See, e.g.,
			V. Gerold and J. Kern, Acta Metall.\
			{\bf 35}, 393 (1987);
			W. Schweika and H. -G. Haubold,
			Phys.\ Rev.\ B {\bf 37}, 9240 (1988);
			L. Reinhard, B. Sch\"onfeld, G. Kostorz,
			and W. B\"uhrer,
			Phys.\ Rev.\ B {\bf 44}, 1727 (1990);
			L. Reinhard, J. L. Robertson, S. C. Moss,
			G. E. Ice, P. Zschack, and C. J. Sparks,
			Phys.\ Rev.\ B {\bf 45}, 2662 (1992).

\bibitem{emf}		M. Bienzle, T. Oishi, and F. Sommer, J. of Alloys and
     			Compounds {\bf 220}, 182 (1995). 

\bibitem{Pasturel}	T. Deutsch and A. Pasturel, in {\em Stability of 
			Materials}, edited by A. Gonis, P. Turchi, and 
			J. Kudrnovsky, NATO-ASI Series (Plenum, 1996). 

\bibitem{Amador}	C. Amador and G. Bozzolo, Phys.\ Rev.\ B {\bf 49}, 
			956 (1994). 

\bibitem{Colinet}	C. Colinet, J. Eymery, A. Pasturel, A. T. Paxton, and
     			M. van Schilfgaarde, J. Phys.: Condens. Matter {\bf 6}, 
			L47 (1994). 

\bibitem{Morgan}	D. Morgan and D. de Fontaine (private communication). 

\bibitem{Mohri}		S. Takizawa, K. Terakura, and T. Mohri, Phys.\ Rev.\ B
     			{\bf 39}, 5792 (1989). 

\bibitem{Asta}		M. Asta and S. M. Foiles, Phys.\ Rev.\ B {\bf 53}, 
			2389 (1996). 

\bibitem{Finel}		R. Tetot and A. Finel, in {\em Stability of Materials}, 
			edited by A. Gonis, P. Turchi, and 
			J. Kudrnovsky, NATO-ASI Series (Plenum, 1996). 

\bibitem{Eymery}	J. Eymery, F. Lancon, and L. Billard,
			J. Phys. I France {\bf 3}, 787 (1993). 

\bibitem{Connolly83}    J. W. D. Connolly and A. R. Williams, Phys.\
                       Rev.\ B {\bf 27}, 5169 (1983).

\bibitem{deFontaine79}  D. de Fontaine, Solid State Phys.\
			{\bf 34}, 73 (1979).

\bibitem{Zunger94}      A recent review is given in
			A. Zunger, in {\em Statics and Dynamics of
			Alloy Phase Transformations}, edited by
			P. E. A. Turchi and A. Gonis, NATO ASI Series
			(Plenum, New York, 1994) p. 361.

\bibitem{Wolverton94}   For a list of many cases in which ASA and full-potential
			formation energies significantly disagree, see
			Table I in C. Wolverton and A. Zunger, Phys\ Rev.\ B
			{\bf 50}, 10548 (1994).

\bibitem{elastic}	D. M. Wood and A. Zunger, Phys.\ Rev.\ B {\bf 40}, 4062 (1989).

\bibitem{Zungerepi}     A. Zunger, in {\em Handbook of Crystal Growth},
			Vol. 3, , D. T. J. Hurle, ed., (Elsevier, 1994).

\bibitem{Bernard94}	J. E. Bernard and A. Zunger, 
	                Appl.\ Phys.\ Lett.\, {\bf 65}, 165 (1994).


\bibitem{note.3}	The Ni-Au system is especially difficult for
			the EAM.  Similar comparisons between EAM and
			LDA for other systems (e.g., Cu-Ag) have yielded
			EAM results significantly closer to LDA. 
			(M. Asta, private communication).

\bibitem{singh}         D. J. Singh, {\em Planewaves, Pseudopotentials,
			and the LAPW Method}, (Kluwer, Boston, 1994).

\bibitem{note}		Generally, it was found that relaxing the cell-internal
			degrees of freedom provided much more energy lowering
			(by roughly a factor of 10) than the energy lowering
			of cell-external coordinates.  For some low
			symmetry monoclinic structures relaxation of the
			length of the unit cell vectors provided an
			insignificant amount of energy lowering, and thus
			the energy lowering associated with the variation
			of the 
			angle of the unit cell vectors was neglected.

\bibitem{wigner}        E. Wigner, Phys.\ Rev .\ {\bf 46}, 1002 (1934).

\bibitem{froyen}        S. Froyen, Phys.\ Rev.\ B {\bf 39},
			3168 (1989)

\bibitem{SQS}           A. Zunger, S.-H. Wei, L. G. Ferreira, and
			J. E. Bernard, Phys.\ Rev.\ Lett.\ {\bf 65}, 
			352 (1990). 

\bibitem{Laks92}        D. B. Laks, L. G. Ferreira, S. Froyen, and
			A. Zunger, Phys.\ Rev.\ B {\bf 46}, 12587 (1992).

\bibitem{Ozolins97}	V. Ozoli\c{n}\v{s}, C. Wolverton, and A. Zunger (to be
			published).

\bibitem{note.2} 	Some of the previous calculations (``f'',
			``i'',``j'' of Table \protect\ref{comp.niau}) 
			estimated the effects of vibrations
			on the phase diagram, either using a simple
			Debye model (``f'') with LDA bulk modulus
			calculations or continuous-space
			Monte Carlo simulations (``i'',``j'') using
			the elastic response of an empirical potential.

\bibitem{Wolverton97}	C. Wolverton, A. Zunger, and B. Schonfeld,
			Solid State Commun.\ {\bf 101}, 519 (1997).






\end{references}
\end{document}